\documentclass[prb,nofootinbib,twocolumn,superscriptaddress]{revtex4} 

\usepackage{graphicx}
\usepackage{dcolumn}
\usepackage{bm}
\usepackage{threeparttable}
\usepackage{times}
\usepackage{mathptmx}
\usepackage{lscape}
\usepackage{natbib}
\usepackage{amsmath}
\usepackage{amssymb}
\usepackage{braket}
\usepackage{comment}
\usepackage{color}

\def\degree{\kern-.2em\r{}\kern-.3em}

\begin{document}


\title{  Special Microscopic-states-basis Formulation of Macroscopic Structure \\for Thermodynamic Systems   }

\author{Koretaka Yuge}
\affiliation{
Department of Materials Science and Engineering,  Kyoto University, Sakyo, Kyoto 606-8501, Japan\\
}%

\author{Shouno Ohta}
\affiliation{
Department of Materials Science and Engineering,  Kyoto University, Sakyo, Kyoto 606-8501, Japan\\
}%

\begin{abstract}
{
For classical system under constant composition, macroscopic structure in thermodynamically equilibrium state can be determined through the so-called canonical average, including sum over possible microscopic states on phase space. Although a set of microscopic structure dominantly contributing to equilibrium properties should depend on temperature and many-body interactions, we recently clarify that at high temperature, they are universally characterized by a \textit{single} special microscopic state (which we call projection state: PS), whose structure can be known a priori without any thermodynamic information. Here we extend this approach to find additional special microscopic states, enabling us to characterize equilibrium structures for lower-temperature region above transition temperature. 
The concept of our approach will lead to a new paradigm; the formulation of macroscopic properties by special microscopic states basis. 
  }
\end{abstract}


\maketitle

\section{Introduction}
When we predict macroscopic structure in thermodynamically equilibrium state for classical systems from microscopic interactions, statistical mechanics provides bridge between macroscopic and microscopic world, the so-called canonical average:
\begin{eqnarray}
\label{eq:q}
Q_{r}\left(T\right) = Z^{-1}\sum_{d}q_{r}^{\left(d\right)}\exp\left(-\beta E^{\left(d\right)}\right),
\end{eqnarray}
where $\beta=\left(k_{\textrm{B}}T\right)^{-1}$ represents inverse temperature, $\left\{q_r\right\}$ denotes a set of prepared coordination, and summation is taken over possible states on phase (or configuration)  space.
From Eq.~(\ref{eq:q}), it is clear that a set of microscopic structure, dominantly contributing to the expectation value $Q_r\left(\beta\right)$, should in principle depend both on temperature and on energy (i.e., microscopic interactions). 
In other words, we cannot \textit{a priori} know which microscopic structures dominate macroscopic properties unless thermodynamic information of temperature or energy is explicitly given. 
Therefore, several techniques have been developed including Metropolis algorism, entropic sampling and Wang-Landau sampling in order to effectively sample important microscopic states for macroscopic properties.\cite{mc1,mc2,mc3,wl}

Despite these facts, we recently find that for classical discrete systems under typical periodic lattices, Eq.~(\ref{eq:q}) can be approximately rewritten as\cite{spe,em1,em3,cm,em2} 
\begin{eqnarray}
\label{eq:qp}
Q_{r}\left(T\right) \simeq g_{1} - \sqrt{\frac{\pi}{2}}g_{2} \cdot \beta{\hat{E_{r}}},
\end{eqnarray}
where $g_1$ and $g_2$ are constant depending only on the type of lattice, and $\hat{E_{r}}$ denotes energy for special microscopic state (\textit{projection state}: PS), whose microscopic structure depends only on lattice.
Eq.~(\ref{eq:qp}) directly tells us that we can \textit{a priori} know a single microscopic state dominantly contributing to $Q_r\left(T\right)$, without requiring any thermodynamic information.
Our derivation relies only on the fact that configurational density of states (CDOS), \textit{before} applying microscopic interactions on the system, is well characterized by a mutidimensional
gaussian distribution at thermodynamic limit: Eq.~(\ref{eq:qp}) becomes exact when CDOS exacly matches gaussian. 
We have confirmed that such character of CDOS holds for representative 3-dimensional lattices, by employing numerical simulation, random matrix theory with gaussian orthogonal ensemble, and by providing explicit formulation of all even-order moments of CDOS under constant composition. 
Althogh the derived expression of Eq.~(\ref{eq:qp}) is universal for any given coordination, deviation of the predicted value of $Q_r$ from typical thermodynamic simulation occurs especially at low-temperature region near and below order-disorder transition temperature. 
The deviation certainly comes from deviation in practical CDOS from ideal gaussian distribution, especially from neglecting contribution for odd-order generalized moment of CDOS. 

Here we extend our previous approach to further include contribution from odd-order moments, finding additional special microscopic states contributing to $Q_r\left(T\right)$ at lower temperature region. 
We demonstrate that generalization of the present extention leads to a new paradigm, i.e., the formulation of macroscopic properties by special microscopic states basis. Details are shown below.

\section{Derivation and Discussions}
Since landscape of multidimensional gaussian can be completely specified by the corresponding covariance matrix, it is clear that previously-found PS includes information about even-order genelarized moment of CDOS, while that about odd-order moments are totally neglected. 
Therefore, to improve our approach to figure out additional special microscopic states characterizing macroscopic properties, it is a natural start to perform Taylor's expansion of canonical average to explicitly include contribution from odd-order moments, leading to
\begin{eqnarray}
\label{eq:q3}
Q_{r}\left(T\right) \simeq g_{1} - \sqrt{\frac{\pi}{2}}g_{2} \cdot \beta{\hat{E_{r}}} + \frac{\beta^2}{2}\sum_{j,k}\Braket{q_r q_j q_k}\Braket{E|q_j}\Braket{E|q_k}, \nonumber \\
\quad
\end{eqnarray}
where $\Braket{q_r q_j q_k}$ denotes third-order moment of CDOS measured from its center of gravity, and $\Braket{\quad|\quad}$ denotes inner product, i.e., trace over possible states on configuration space. 
Note that although Eq.~(\ref{eq:q3}) appears to merely be moment-expansion of canonical average up to third-order moment, the equation automatically includes partial information about all even-order moments: This is because practical CDOS, again, can be well characterized by multidimensional gaussian for large systems. 
In analogy to our previous study, we should find out additional special microscopic state(s) to determine the value of summation in the third term of right-hand side of Eq.~(\ref{eq:q3}). 
Since unit of the considered term is squared energy, energy for a single microscopic state cannot determine its value. 
When we try to rewrite the summation by product of energies of two microscopic states, it leads to
\begin{widetext}
\begin{eqnarray}
\label{eq:qqq}
\sum_{j,k}\Braket{q_r q_j q_k}\Braket{E|q_j}\Braket{E|q_k} = \left(\sum_p \Braket{E|q_p}q_p^{\left(1\right)}  \right)\cdot \left(\sum_s \Braket{E|q_s}q_s^{\left(2\right)}  \right). \nonumber \\
\end{eqnarray}
\end{widetext}
For $f$-dimensional configuration space considered, to determine the values of a set of $\left\{q_p^{\left(1\right)}\right\}$ and of $\left\{q_s^{\left(2\right)}\right\}$, we have $2f$ unknown numbers, while Eq.~(\ref{eq:qqq}) leads to $f^2/2$ independent terms included in the third-order moments. Furthermore, our recent study reveals that third-order moments for any combination of three figures $r$, $j$ and $k$ all have the same system-size dependence, which directly means that we cannot generally rewrite the third term of Eq.~(\ref{eq:q3}) by the right-hand side of Eq.~(\ref{eq:qqq}). 
To overcome the problem, we first rewrite left-hand side of Eq.~(\ref{eq:qqq}) as explicit quadratic form by corresponding matrix, namely
\begin{eqnarray}
\sum_{j,k}\Braket{q_r q_j q_k}\Braket{E|q_j}\Braket{E|q_k} = {}^t\!X \mathbf{A^{\left(r\right)}} X,
\end{eqnarray}
where $X$ and $\mathbf{A^{\left(r\right)}}$ are $f$-dimensional vector and $f\times f$ symmetric matrix, respectively given by
\begin{eqnarray}
{}^t\!X &=& \left(\Braket{E|q_1},\cdots,\Braket{E|q_f}\right) \nonumber \\
A_{ij}^{\left(r\right)} &=& \Braket{q_r q_i q_j}.
\end{eqnarray}
Therefore, in order to decompose the summation into products of energy for a certain set of microscopic state, we perform singular value decomposition (SVD) of matrix $\mathbf{A^{\left(r\right)}}=\mathbf{UDV}^T$, leading to
\begin{widetext}
\begin{eqnarray}
\sum_{j,k}\Braket{q_r q_j q_k}\Braket{E|q_j}\Braket{E|q_k} = \sum_{k=1}^f {}^t\!X \lambda_k \left(\mathbf{U}_k \otimes \mathbf{V}_k\right) X =    \sum_{k=1}^{f} \left\{ {}^t\!X\cdot \left(\lambda_k^{\frac{1}{2}} \mathbf{U}_k \right)\right\} \left\{ \left(\lambda_k^{\frac{1}{2}} \mathbf{V}^T_k\right)\cdot X    \right\}, 
\end{eqnarray}
\end{widetext}
where $\lambda_k$ denote $k$-th singular value of $\mathbf{A}$, and $\mathbf{U}_k$ and $\mathbf{V}_k$ are $k$-th column of matrix $\mathbf{U}$ and $\mathbf{V}$, respectively.
Since only $X$, corresponding to inner products, depends on many-body interaction in the system, and the all values in $\lambda_k^{\frac{1}{2}} \mathbf{U}_k$ and $\lambda_k^{\frac{1}{2}} \mathbf{V}^T_k$ can be known \textit{a priori} without any thermodynamic information, we can approximate the summation by the product of energies for two \textit{specially selected} microscopic states,
\begin{eqnarray}
\sum_{j,k}\Braket{q_r q_j q_k}\Braket{E|q_j}\Braket{E|q_k} \simeq \hat{E}_r^{\left(3,1\right)}\cdot \hat{E}_r^{\left(3,2\right)},
\end{eqnarray}
whose microscopic structures are respectively given by
\begin{eqnarray}
\label{eq:sp}
&&\mathrm{str1:\quad}\left\{\lambda^{\frac{1}{2}}_m U_{1m},\cdots, \lambda^{\frac{1}{2}}_m U_{fm}\right\} \nonumber \\
&&\mathrm{str2:\quad}\left\{\lambda^{\frac{1}{2}}_m V_{1m},\cdots, \lambda^{\frac{1}{2}}_m V_{fm}\right\},
\end{eqnarray}
where $\lambda_m$ denotes the largest singular value. The important point is, again, structure of two special microscopic states given in Eq.~(\ref{eq:sp}) can be known \textit{a priori} without any thermodynamic information. 
We finally note that extention of the present approach to figure out additional special microscopic states for higher-order moments is non-trivial: This is because higher-order moment requires at least 3-order tensor, which cannot be decomposed into product (not Kronecker product) of multiple vectors, which should be considered in our future study.

\section{Conclusions}
We find a new set of special microscopic states to characterize macroscopic structure in thermodynamically equilibrium state especially at low temperatures near order-disorder transition temperature, where the structure of the microscopic states can be known \textit{a priori} without any thermodynamic simlumation. The derivation is based on including information about odd-order generalized moments of CDOS, which has been totally neglected in a special state we previously found. Since the present derivation can be applied to any number of components at any composition, application of the microscopic states to DFT calculation will make significant advances in clarify microscopic structure for e.q., high-entropy alloy where number of constituents typically exceeds four or five.

\section{Acknowledgement}
This work was supported by a Grant-in-Aid for Scientific Research (16K06704) from the MEXT of Japan, Research Grant from Hitachi Metals$\cdot$Materials Science Foundation, and Advanced Low Carbon Technology Research and Development Program of the Japan Science and Technology Agency (JST).

\end{document}